# A proof-of-principle demonstration of quantum microwave photonics

YAQING JIN,[1,3] YE YANG,[2,6] HUIBO HONG,[1,3] XIAO XIANG,[1,3] RUNAI QUAN,[1,3] TAO LIU,[1,3] SHOUGANG ZHANG,[1,3] NINGHUA ZHU,[2,4,5] MING LI,[2,4,5,#] AND RUIFANG DONG [1,3,*]

[1] *Key Laboratory of Time and Frequency Primary Standards, National Time Service Center, Chinese Academy of Sciences, , Xi'an 710600, China*
[2] *State Key Laboratory on Integrated Optoelectronics, Institute of Semiconductors, Chinese Academy of Sciences, Beijing, 100083, China*
[3] *School of Astronomy and Space Science, University of Chinese Academy of Sciences, Beijing 100049, China*
[4] *School of Electronic, Electrical and Communication Engineering, University of Chinese Academy of Sciences, Beijing 100049, China*
[5] *Center of Materials Science and Optoelectronics Engineering, University of Chinese Academy of Sciences, Beijing 100190, China*
[6] *The 29th Research Institute of China Electronics Technology Group Corporation, Chengdu 610029, China*
*Corresponding author [#]ml@semi.ac.cn, *dongruifang@ntsc.ac.cn*

**Abstract:** With the rapid development of microwave photonics, which has expanded to numerous applications of commercial importance, eliminating the emerging bottlenecks becomes of vital importance. For example, as the main branch of microwave photonics, radio-over-fiber technology provides high bandwidth, low-loss, and long-distance propagation capability, facilitating wide applications ranging from telecommunication to wireless networks. With ultrashort pulses as the optical carrier, huge capacity is further endowed. However, the wide bandwidth of ultrashort pulses results in the severe vulnerability of high-frequency RF signals to fiber dispersion. With a time-energy entangled biphoton source as the optical carrier and combined with the single-photon detection technique, a quantum microwave photonics method is proposed and demonstrated experimentally. The results show that it not only realizes unprecedented nonlocal RF signal modulation with strong resistance to the dispersion associated with ultrashort pulse carriers but provides an alternative mechanism to effectively distill the RF signal out from the dispersion. Furthermore, the spurious-free dynamic range of both the nonlocally modulated and distilled RF signals has been significantly improved. With the ultra-weak detection and high-speed processing advantages endowed by the low-timing-jitter single-photon detection, the quantum microwave photonics method opens up new possibilities in modern communication and networks.

## 1. Introduction

Microwave photonics (MWP), which uses photonic techniques to deal with microwave signals, has extended numerous applications covering broadband wireless access networks, sensor networks, radar, satellite communications, and warfare systems [1, 2]. At the same time, the bottlenecks of the MWP have also emerged, which place restrictions on its usability. For example, as the main branch of MWP, radio-over-fiber (RoF) uses optical fibers for transmitting microwave signals to a distant receiver. Benefitted from the advantages of high bandwidth and low loss of optical fibers, RoF shows excellent capabilities of high speed and long-distance propagation [3-7]. To support the rapidly increased demand for the high capacity, the ultrafast optical division multiplexing (OTDM) technique based on ultrashort pulses becomes a requisite in RoF technology[1, 8]. However, due to the large spectral width of ultrashort pulses, the transmitted microwave signal is severely influenced by the fiber chromatic dispersion. In addition, suppression of the electrical distortion from harmonic and intermodulation is also crucial for the performance of MWP.

Over the past decades, quantum correlations between photon pairs have provided diversified access to surpass classical capabilities in various applications, including communication, simulation, computation, etc.[6, 9]. For instance, by using the spatial correlation between the photon pairs, the quantum imaging system[10-12] enables the nonlocal obtaining of the image information of the object that is hardly reachable via direct detection with the fewest number photons and sub-diffraction resolution[13]. With these virtues, quantum imaging has been extensively developed[14-17] and extended to abundant applications such as quantum-secured imaging[18], quantum illumination[19], and wide-field microscope[20]. The temporal correlation between the photon pairs has also been effectively applied in quantum clock synchronization[21-24], quantum key distribution[25, 26], and quantum spectroscopy[27, 28]. Recently, nonlocal temporal shaping of the photons has been proposed[29, 30]. Therefore, it is expected to realize the nonlocal RF modulation and quantum enhancement in MWP technology.

In this paper, we report a quantum microwave photonics (QMWP) scheme with the time-energy entangled biphoton source as the optical carrier and the single-photon detection. Based on this scheme, the RF signal modulated on the idler photons can be nonlocally mapped onto the signal photons, which shows improved spurious-free dynamic range (SFDR) in terms of second harmonic distortion and strong resistance to the dispersion impact associated with the ultrashort pulse carrier. Furthermore, by applying coincidence-based selection to the idler photons, SFDR improvement of the direct RF modulation and anti-dispersion capability of the RoF signal has also been achieved. Combined with the advantages of ultra-weak detection and high-speed processing rendered by the single-photon MWP (SP-MWP) [31]method, the QMWP provides unprecedented feasibilities to the MWP technology.

## 2. Principle model

The principle diagram of classical MWP in the RoF link is plotted in Fig. 1(a). Given a broadband optical source as the optical carrier, its temporal amplitude function can be written as

$$E(t) \sim \exp(-i\omega_0 t)\exp\left(-\frac{t^2}{\tau_p^2}\right), \tag{1}$$

where $\omega_0$ and $\tau_p$ denote the center frequency and temporal duration of the optical source respectively. When a radio frequency (RF) of $\omega_{RF}$ is intensity-modulated onto the optical source, whose impulse response function is described by

$$h(t) \propto 1 + \cos(\omega_{RF} t), \tag{2}$$

the temporal amplitude function of the modulated optical signal is translated into

$$E'(t) = h(t)E(t), \tag{3}$$

To evaluate the propagation characteristic of the modulated broadband optical signal through fiber, the Fourier transform is applied onto Eq. (3) to give the spectral wave function as

$$\tilde{E}'(\omega) \sim \exp\left[-\frac{\tau_p^2}{4}(\omega-\omega_0)^2\right] + \frac{1}{2}\left\{\exp\left[-\frac{\tau_p^2}{4}(\omega+\omega_{RF}-\omega_0)^2\right] + \exp\left[-\frac{\tau_p^2}{4}(\omega-\omega_{RF}-\omega_0)^2\right]\right\}, \tag{4}$$

The transfer function of the fiber is given by

$$G(\omega) = \exp\left\{-i\left[\beta_1(\omega-\omega_0) + \beta_2(\omega-\omega_0)^2\right]\right\}, \tag{5}$$

where $\beta_1$ and $\beta_2$ denote the group delay and dispersion introduced by the fiber respectively. After experiencing the fiber propagation, the spectral wave function given by Eq. (4) is transferred to

$$\tilde{E}''(\omega) = G(\omega)\tilde{E}'(\omega), \qquad (6)$$

Its temporal amplitude function is derived by the inverse Fourier transformation of $\tilde{E}''(\omega)$ as

$$E''(t) \sim \exp\left[-\frac{(t+\beta_1)^2}{\tau_p^2 \beta_2^2}\right] + \frac{1}{2}\left\{\begin{array}{l}\exp[i\omega_{RF}(t+\beta_1-\beta_2\omega_{RF})]\exp\left[\frac{-(t+\beta_1-2\beta_2\omega_{RF})^2}{\tau_p^2\beta_2^2}\right] \\ +\exp[i\omega_{RF}(t+\beta_1+\beta_2\omega_{RF})]\exp\left[\frac{-(t+\beta+2\beta_2\omega_{RF})^2}{\tau_p^2\beta_2^2}\right]\end{array}\right\}, \qquad (7)$$

As seen from Eq. (7), the fiber dispersion not only leads to the pulse duration broadening and introduces different phase and temporal shifts to the two modulation-associated terms. The different phase shifts give rise to the well-known frequency-dependent fading to the RF modulation[1], while the different temporal shifts introduce a more severe dispersion distortion to the transmitted RF signal.

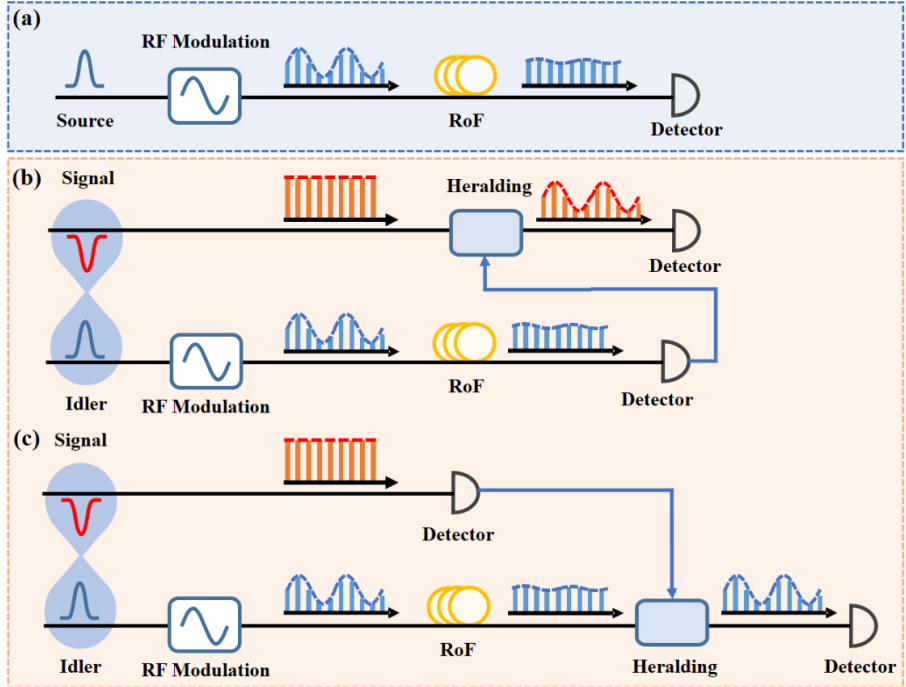

Fig. 1. Principle diagrams of the classical MWP in the RoF link with classical broadband optical source (a) and the quantum MWP in the RoF link with time-energy entangled biphoton source: (b) depicts the nonlocally acquired RF from the heralded signal photons, (c) depicts the RF distillation from dispersion distortion from the heralded idler photons.

In comparison, the QMWP in the ROF link is depicted in Fig. 1(b) and (c). According to quantum theory, the temporal wave function of the time-energy entangled biphoton source can be given by

$$\Psi(t_1, t_2) \propto \exp\left[-i(\omega_{s,0}t_1 + \omega_{i,0}t_2) - \frac{(t_1-t_2)^2}{\tau_c^2}\right], \tag{8}$$

where $\omega_{s(i),0}$ denotes the center angular frequency of the signal (idler) photons, $\tau_c$ is the temporal coincidence width of the biphotons. The temporal coordinates of the signal and idler photons are expressed by $t_1$ and $t_2$, respectively. The square modulo of the temporal wave function defines the second-order Glauber correlation function ($G^{(2)}$), i.e., $G^{(2)} \equiv |\Psi(t_1,t_2)|^2$.

When the RF signal with $\omega_{RF}$ is intensity modulated onto the idler photons, the two-photon state function is transferred to

$$\Psi'(t_1,t_2) = h(t_2)\Psi(t_1,t_2), \tag{9}$$

Under the condition that the idler photons are subsequently traveled through the long-distance optical fiber, whose transfer function is given by Eq. (4), the temporal wave function of the two-photon state after the modulation and dispersion can then be given by

$$\Psi''(t_1,t_2) \sim \exp\left[\frac{-(t_1-t_2-\beta_1)^2}{\tau_c^2\beta_2^2}\right] + \frac{1}{2}\left\{\begin{array}{l} \exp\left[-i\omega_{RF}(t_2+\beta_1+\beta_2\omega_{RF})\right]\exp\left[\frac{-(t_1-t_2-\beta_1-2\beta_2\omega_{RF})^2}{\tau_c^2\beta_2^2}\right] \\ + \exp\left[i\omega_{RF}(t_2+\beta_1-\beta_2\omega_{RF})\right]\exp\left[\frac{-(t_1-t_2-\beta_1+2\beta_2\omega_{RF})^2}{\tau_c^2\beta_2^2}\right] \end{array}\right\}, \tag{10}$$

One can notice that, Eq. (10) is similar with Eq. (7) except that the evolution of $t$ is replaced by $t_2 - t_1$. As shown in Fig. 1(b), conditioned by the second-order Glauber correlation between the signal and idler photons, the heralding of the signal photons can be operated. The conditional selection is achieved by applying a coincidence window function of $S(t_1 - t_2 - \beta_1)$ onto $\Psi''(t_1,t_2)$, which is given by a rectangular function

$$S(t_1-t_2-\beta_1) = \begin{cases} 1, |t_1-t_2-\beta_1| \leq \frac{\tau}{2} \\ 0, |t_1-t_2-\beta_1| > \frac{\tau}{2} \end{cases}, \tag{11}$$

Here $\tau$ denotes the window width. The temporal wave function of the heralded signal photons is then derived after integration over $t_2$ and can be written as

$$\varphi_s''(t_1) \sim \int dt_2 S(t_1-t_2-\beta_1)\Psi''(t_1,t_2) \propto \text{erf}\left(\frac{1}{\tau_c\beta_2}\right) + C_1\exp(-i\beta_2\omega_{RF}^2)\cos(\omega_{RF}t_1) \tag{12}$$
$$\approx 1 + C_1\exp(-i\beta_2\omega_{RF}^2)\cos(\omega_{RF}t_1)$$

Here $\text{erf}\left(\frac{1}{\tau_c\beta_2}\right) \approx 1$ as $\tau_c\beta_2 \ll 1$. $C_1 = \frac{1}{2}\left(\text{erf}\left(\frac{\tau+4\beta_2\omega_{RF}-i\omega_{RF}\tau_c^2\beta_2^2}{2\tau_c\beta_2}\right) + \text{erf}\left(\frac{\tau-4\beta_2\omega_{RF}+i\omega_{RF}\tau_c^2\beta_2^2}{2\tau_c\beta_2}\right)\right)$. As long as $\tau \gg 4\beta_2\omega_{RF} \gg \omega_{RF}\tau_c^2\beta_2^2$ is satisfied, $C_1 \approx \text{erf}\left(\frac{\tau}{2\tau_c\beta_2}\right)$. When $\tau \gg 2\tau_c\beta_2$, $C_1 \approx 1$. In this case, Eq. (12) is reduced to

$$\varphi_s''(t_1) \approx 1 + \exp(-i\beta_2\omega_{RF}^2)\cos(\omega_{RF}t_1), \tag{13}$$

which follows the same expression as the classical case with a continuous-wave light being the carrier [1]. In comparison with Eq. (7), the dispersion-induced distortion due to the broad bandwidth of the pulse carrier has no influence on the RF signal nonlocally acquired from the heralded signal photons. Therefore, with the time-energy entangled biphoton source and coincidence heralding, the QMWP is capable of nonlocally acquiring the RF signal with excellent anti-dispersion property.

As depicted in Fig. 1(c), conditioned by the second-order Glauber correlation between the signal and idler photons, the heralding of the idler photons can also be operated. The temporal wave function of the heralded idler photons can be written as

$$\varphi_i''(t_2) \sim \int dt_1 S(t_1 - t_2 - \beta_1) \Psi''(t_1, t_2)$$
$$\propto \operatorname{erf}\left(\frac{1}{\tau_c \beta_2}\right) + C_2 \exp\left(-i\beta_2 \omega_{RF}^2\right) \cos\left(\omega_{RF} t_2 + \omega_{RF} \beta_1\right), \quad (14)$$
$$\approx 1 + C_2 \exp\left(-i\beta_2 \omega_{RF}^2\right) \cos\left(\omega_{RF}(t_2 + \beta_1)\right)$$

We can see that Eq. (14) has a similar form with Eq. (12) except a different factor of $C_2$, which is given by $C_2 = \frac{1}{2}\left(\operatorname{erf}\left(\frac{\tau + 4\beta_2 \omega_{RF}}{2\tau_c \beta_2}\right) - \operatorname{erf}\left(\frac{\tau - 4\beta_2 \omega_{RF}}{2\tau_c \beta_2}\right)\right) < 1$. By choosing $\tau$ to maximize $C_2$, the RF signal on the idler photons can be distilled from the dispersion distortion.

3. **Experimental Setup**

The experimental setup of the QMWP in the RoF link is shown in Fig. 2. Figure 2(a) depicts the setup for the generation of the utilized time-energy entangled biphoton source. Via the spontaneous parametric down-conversion process (SPDC), the time-energy entangled biphoton source at 1560 nm is generated from a 10 mm-long, type-II quasi-phase-matched periodically poled lithium niobate (PPLN, HC Photonics) waveguide with a period of 8.3 μm. The pump laser is a 780 nm DBR laser diode (Photodigm) with a linewidth of 2 MHz[32]. As the PPLN waveguide is integrated with the polarization-maintaining single-mode fiber (PM-SMF), the produced photon pairs can be directly coupled into the fiber channels. By inserting a customized WDM (PMFWDM-5678-222, Optizone) at the PPLN output, the residual pump can be effectively removed. Afterward, the orthogonally polarized signal and idler photons are spatially separated by a fiber polarization beam splitter (FPBS) for subsequent utilization. Figure 2(b) shows the RoF link. On the idler photon path, intensity modulation is applied by employing an external Mach–Zehnder modulator (MZM, PowerBit™ F10-0, Oclaro). The modulator is driven by a sinusoidal RF signal from a signal generator (SG382, Stanford Research). The fiber-Bragg-grating-based dispersion compensation module (DCM, Proximion AB) is used to simulate the dispersion in long-distance fibers. In our experiment, different DCMs with the GVD ranging from 165 ps/nm to 826 ps/nm are applied. Figure 2(c) plots the setup for measurement of the RF signal based on the QMWP method, which utilizes the coincidence-based post-selection to verify the heralding.

In our experiment, both the signal and idler photons are detected by the low-jitter superconductive nanowire single-photon detectors (SNSPDs, Photec). The full-width-at-half-maxima (FWHM) of the time jitter of the SNSPD is about 50 ps[33]. The two SNSPD outputs are then delivered to different input ports (CH1 and CH2) of the TCSPC module (PicoQuant Hydraharp 400), which is operated in the Time Tagged Time-Resolved (TTTR) T3 mode. In this mode, the TCSPC records the arrival times of the photon events for each input port, the measurement time for the photon event recording is set as 180 s. The sampling frequency is inversely proportional to the time-bin resolution of the TCSPC, which is set as 8 ps and the corresponding sampling frequency is $F_s = 125$ GHz. With the 10 MHz time base signal from the SG382 as the sync input for the TCSPC, one can extract the photon waveforms of the signal and idler as a function of the periodic sync input from the individual photon records, which show the RF modulation carried by the photons. The maximum range of the photon waveform is determined by the sync signal period, which is 100 ns. It should be pointed out that, to conveniently keep the phase stabilization between the modulated RF signal and the sync signal, we use the time base output of the SG382 as the sync signal. Using another time base delivered both to the sync input of the TCSPC and the time base input of the SG382 is also practicable. By only looking at the idler photon path, this setup is equivalent to the SP-MWP system presented in [31]. From the above individually acquired photon events by the TCSPC, the second-order cross-correlation between the signal and idler can be further analyzed and used to perform the coincidence-based post-selection on either the signal or the idler photons. To realize the post-selection, the coincidence histogram needs to be first obtained. Based on this histogram, the paired photon can be identified and selected within a certain coincidence window. Via this program, the selected photon events both for signal and idler as a function of the sync event can be obtained to reveal the heralded photon waveforms over time.

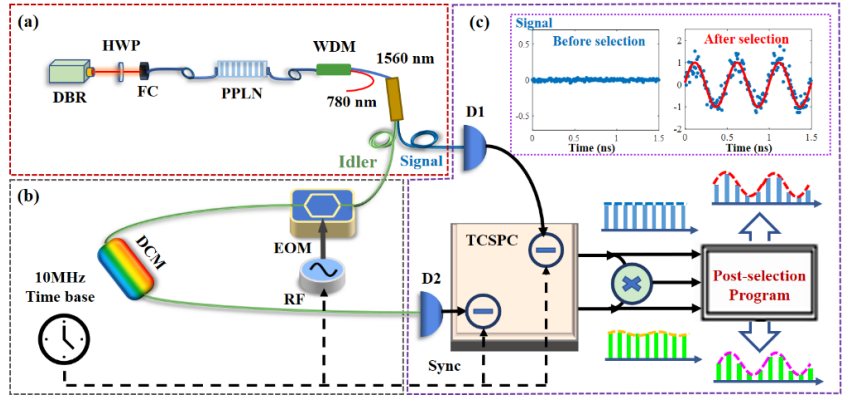

Fig. 2. Experimental setup of the QMWP in a simulated RoF link. (a) depicts the setup for the generation of the utilized time-energy entangled biphoton source. (b) plots the RoF link. The idler photons of the time-energy entangled biphoton source are modulated by a high-speed RF signal ($\omega_{RF}$) via the EOM and then transmitted through the DCM to simulate the RoF. (c) shows the setup for measurement and nonlocal recovery of the RF signal based on the QMWP method. The SNSPDs D1 and D2 are used to detect the arrived signal and idler photons. The TCSPC in the TTTR mode is used to record the time arrivals of the SNSPDs, and its sync input shares the same 10 MHz time base with the RF signal. The time differences between the individually recorded photon events and the relevant last sync event measure the photon waveforms over time. Based on the auxiliary cross-correlation searching program, the coincidence between the recorded signal and idler photon events is acquired and used for the post-selection. The inset of (c) plots the direct and coincidence-assisted measurement of the signal photon waveforms.

4. **Experimental Results and Analysis**

The nonlocal mapping of the RF modulation on the idler photons to the signal photons is firstly investigated without the DCM in the setup. By setting the RF signal at 2 GHz with a modulation power of 10 dBm, Fig. 3(a) depicts direct measurements of the signal (blue) and idler (green) photon waveforms over time. Note that the recorded counts of the idler in Fig. 3(a) are much fewer than that of the signal photons which is due to the inserted loss by the intensity modulator. One can see the RF modulation from the trace of the idler photons, while that of the signal photons is clean from any modulation. Applying Discrete-Fourier-Transform (DFT) to the waveforms, the corresponding power spectra of the signal (blue) and idler (green) are given in Fig. 3(b). The power ratio of the DFT noise floor to the signal is 55.8 dB. The DFT noise floor is $10\log(N/2)$ dB below the actual noise floor, where $N$ is the number of bins for the DFT. To improve the performance of DFT, $N$ is set to the next power of 2 from the original temporal waveform length[34]. In our experiment, $N = 2^{14}$ and $10\log(N/2) = 39.1$ dB. After correcting the DFT noise floor, the 2 GHz RF with a signal-to-noise ratio (SNR) of 16.7 dB is shown for the idler, while no trivial RF component can be seen from the signal, which shows that the RF modulation can only be observed on the idler path based on the SP-MWP. To showcase the nonlocal mapping, post-selection is subsequently applied to the signal photons.

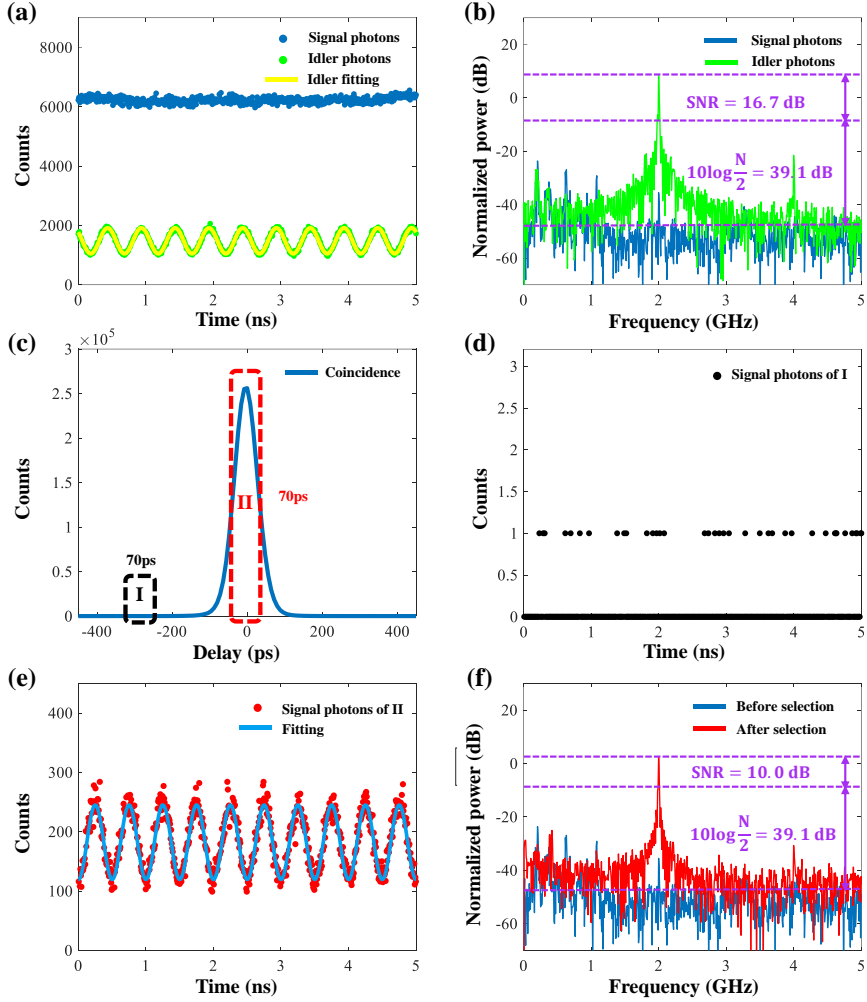

Fig. 3. The experimental results of the nonlocal mapping of the RF signal based on the QMWP scheme, in which the RF signal is set as 2 GHz with a modulation power of 10 dBm and no DCM is put in the setup. (a) The measured signal (blue) and idler (green) photon waveforms over time based on the direct detection. (b) The DFT spectra of the signal (blue) and idler (green) photon waveforms. (c) The acquired coincidence histogram between the signal and idler photons. Under the conditions that the heralding window is far from (I) and at the center of (II) the coincidence histogram, the reconstructed photon waveforms after the post-selection are respectively plotted in (d) and (e). It can be seen that only the signal photons post-selected from the coincidence histogram can output the RF-modulated periodic waveform. The power spectra of the measured signal photon waveforms before (blue) and after (red) the post-selection are shown in (f).

As shown in Fig. 3(c), the coincidence distribution between the signal and idler is found according to the cross-correlation program, which has a FWHM temporal width of 70 ps and is determined by the timing jitters of the SNSPDs. According to the heralding principle, only the photon events within the coincidence distribution range can be treated as the entangled photon pair. To manipulate the post-selection, the heralding time window ($\tau$) that defines the relative time difference between the signal and idler should be set. The recorded signal events whose arrival times fall in the time window are selected out for the RF acquisition. By setting the width of the heralding window to 70 ps, two sets of signal photon events are post-selected when the heralding window is put far from (I) and at the center of (II) the coincidence histogram and used to rebuild the photon waveform over time. The results are shown in Fig. 3(d) and Fig. 3(e), respectively. As seen in Fig. 3(d), only a small number of signal events can be selected since the heralding window I is out of reach of the coincidence range. While in Fig. 3(e), the post-selected signal photons from the coincidence histogram present the periodic waveform featuring the RF modulation. Therefore, the RF modulation on the idler photons is shown nonlocally mapped onto its entangled counterpart (signal photons). Further looking at the spectrum of Fig. 3(e), which is shown in Fig. 3(f), the SNR of 10.0 dB for the RF signal is obtained. Comparing Fig. 3(b) with Fig. 3(f), the SNR of the nonlocally acquired RF signal is degraded by 6.7 dB, which can be mainly attributed to the heralding efficiency as the detected photon counts for the idler is about $1.7 \times 10^7$ while the post-selected photon counts for the signal is $2.0 \times 10^6$.

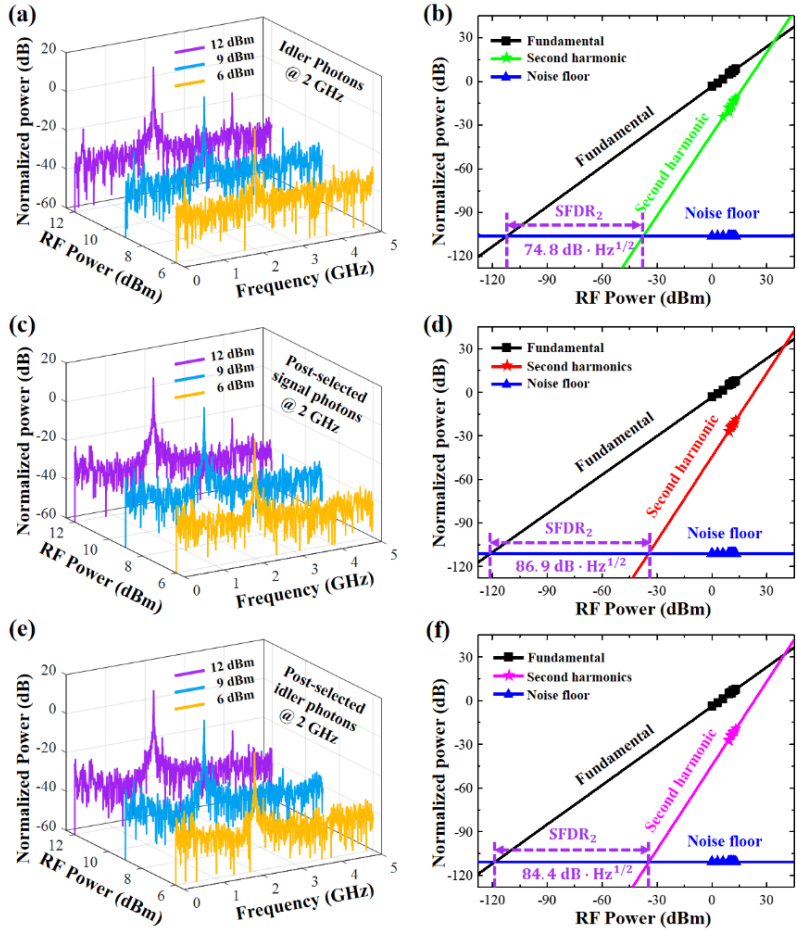

Fig. 4. The DFT spectra of (a) the directly measured idler photon waveforms, (c) the post-selected signal photon waveforms, and (e) the post-selected idler photon waveforms. Different

RF modulation powers are applied for comparing the SFDR regarding the direct detection and post-selected results. The 2 GHz RF signal and noise powers for the fundamental and second harmonic components at different modulation powers are plotted in (b), (d) and (f). By applying linear fitting to these measurements in (b), (d) and (f), the SFDR of the system can be extrapolated.

By comparison between Fig. 3(b) and (f), the nonlocally acquired RF signal by the QMWP method exhibits apparent suppression of the second harmonic distortion. To prove this improvement, the DFT spectra of the directly measured idler photon waveforms (Fig. 4(a)), the post-selected signal photon waveforms (Fig. 4(c)), and the post-selected idler photon waveforms (Fig. 4(e)) are respectively plotted. In these measurements and post-selecting implementations, the photon counts are maintained the same for comparison. Different RF modulation powers are applied for comparing the suppression of the second harmonic distortion regarding the direct detection and post-selected results. From Fig. 4(a) one can see distinct second harmonic distortion as the RF modulation power increases. Meanwhile, the photon spectra of the post-selected signal and idler photons are depicted in Fig. 4(c) and Fig. 4(e), and apparent suppression of the second harmonic distortion can be seen. In addition, the noise floor in Fig. 4(c) and (e) is 4.8 dB and 4.4 dB smaller than that in Fig. 4(a), respectively, and the SNR of the post-selected photons is higher than that of the directly measured photons with the same photon counts. For example, When the modulated signal power is 9 dBm, the SNRs of the selected idler photons and signal photons are 7.9 dB and 7.7 dB, which are 3.8 dB and 3.6 dB larger than that of the unselected photon (4.1 dB). The 2 GHz RF signal and noise power for the fundamental and second harmonic components at different RF powers are plotted in Fig. 4(b), (d) and (f). Considering the spectral resolution of the DFT is $B = F_s/N$ (7.63 MHz in our experiment), which is the RF bandwidth of each DFT noise bin. Therefore, the noise power density $N_{out}$ is $10 \lg(7.63 \times 10^6)$ =-68.83 dB lower than the DFT noise floor. By applying linear fitting to these measurements in (b), (d) and (f), the SFDRs of these three cases, of which the noise power is normalized to $B = 1$ Hz [1], can be extrapolated. For the direct measured result (Fig. 4(b)), the $SFDR_2$ is 74.8 dB $Hz^{1/2}$. In contrast, for the acquired photon signal from the post-selected signal (Fig. 4(d)) and idler photons (Fig. 4(f)), the $SFDR_2$ can be significantly improved to 86.9 dB $Hz^{1/2}$ and 84.4 dB $Hz^{1/2}$ respectively. By increasing the photon counts, the normalized noise floor can be further decreased and the $SFDR_2$ can be improved.

Next, we test the resistance to the dispersion distortion by the QMWP system with the DCM in the setup. Figure 5(a) presents the direct measurement of the signal (blue dots) and idler (green dots) photon waveforms by setting the RF modulation at 2 GHz with a modulation power of 10 dBm and using the DCM with a dispersion of 495 ps/nm. Their corresponding spectra of the temporal distribution of Fig. 5(a) are plotted in Fig. 5(b). Even though the spectral power of the idler photon at 2 GHz is larger than the DFT noise floor, it is still smaller than the actual noise floor, which is 39.1 dB higher than the DFT noise floor. Compared with the results in Fig. 3(a) and (b), the RF modulation on the idler path cannot be detected anymore due to the dispersion. In contrast, we can recover the RF modulation signal by applying coincidence-based post selection on the signal photons. The post-selected photon waveform, whose heralding time window width is 1.3 ns, is shown in Fig. 5(c) by red dots, and its spectrum in the frequency domain is given by the red curve in Fig. 5(d). The spectral power of the selected photon is higher than the actual noise floor, and the SNR is 6.3 dB. The results show that the nonlocally acquired RF signal from the post-selected signal photons can be immune from the dispersion distortion associated with the ultrashort pulse carrier. In the system with a large GVD, the coincidence width is much larger than the timing jitter of the system. To investigate the highest SNR at different heralding time window widths, Figure 5(e) plots the SNR of the nonlocally acquired RF signal as a function of the heralding time window width when the GVD is 495 ps/nm. The highest SNR (6.4 dB) is achieved when the heralding window width is 1.2 ns, slightly smaller than the FWHM of the measured coincidence. This anti-dispersion property is further tested by varying the dispersion from 165 ps/nm to 826 ps/nm, and the SNR of the recovered RF as a function of the dispersion is shown in Fig. 5(f). Compared with the relevant photon counts after post selection, which are also shown in Fig. 5(f) by black curve, the degradation of the SNR agrees well with the post-selected photon counts.

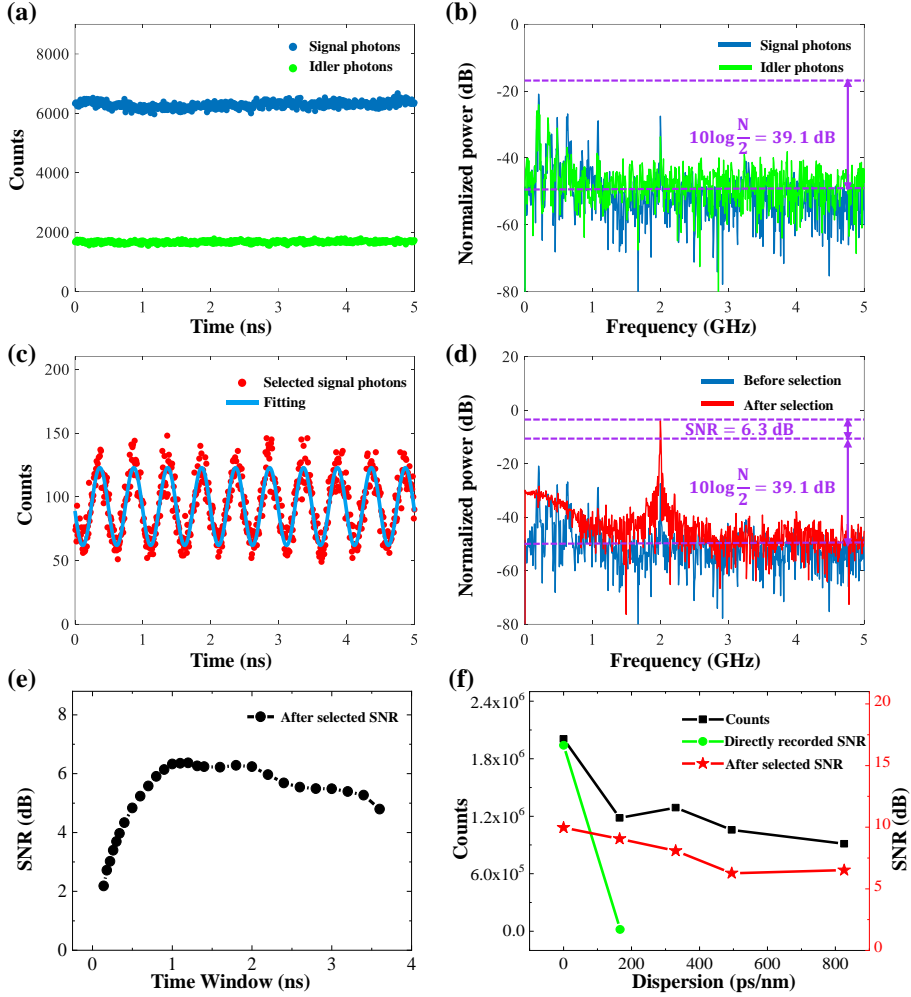

Fig. 5. The experimental results of the nonlocal mapping of the RoF signal based on the QMWP scheme, in which the RF signal is set as 2 GHz with a modulation power of 10 dBm and the utilized DCM has a dispersion of 495 ps/nm. (a) The measured signal (blue) and idler (green) photon waveforms are based on direct detection. (b) The power spectra of the signal (blue) and idler (green) photon waveforms. (c) The reconstructed signal photon waveforms after the post-selection are applied. (d) The power spectra of the measured signal photon waveforms after (red) the post-selection. (e) The plot of the recovered RF signal SNR as a function of the heralding time window width, which shows a maximum when the heralding window width is equal to the FWHM of the measured coincidence. (f) The plot of the recovered RF signal SNR as a function of the dispersion.

Further applying the post-selection onto the idler photons, we have also demonstrated efficient suppression of the dispersion distortion onto the RF signal. Under the same RF modulation and dispersion condition as above, the distilled waveform from the post-selected idler photons (pink dots) is plotted in Fig. 6(a). The corresponding DFT spectra of them are shown in Fig. 6(b), which shows the successful acquisition of a 2 GHz RF signal from the post-selected idler photons with the SNR of 1.9 dB. The achievable SNR of the RF signal as a function of the heralding window width is further plotted in Fig. 6(c), which shows a maximum SNR when the width is chosen as 0.18 ns. Additionally, changing the dispersion from 165 ps/nm to 826 ps/nm, the maximally achievable SNR under different dispersions shows that the optimized width for post-selection is almost constant at 0.18 ns. The achieved maximum SNR as a function of the dispersion is given in Fig. 6(d). Because the coincidence width increases with the growth of the GVD, the photon count in the heralding width decreases dramatically. However, similar to the results shown in Fig. 5(f), the degradation of the SNR coincides with the post-selected photon counts. Furthermore, though the SNR of the distilled RF signal from the idler photons is worse than that from the signal photons, it witnesses successful dispersion distortion suppression that cannot be overcome other than dispersion compensation. By increasing the photon counts, the SNR can be effectively improved.

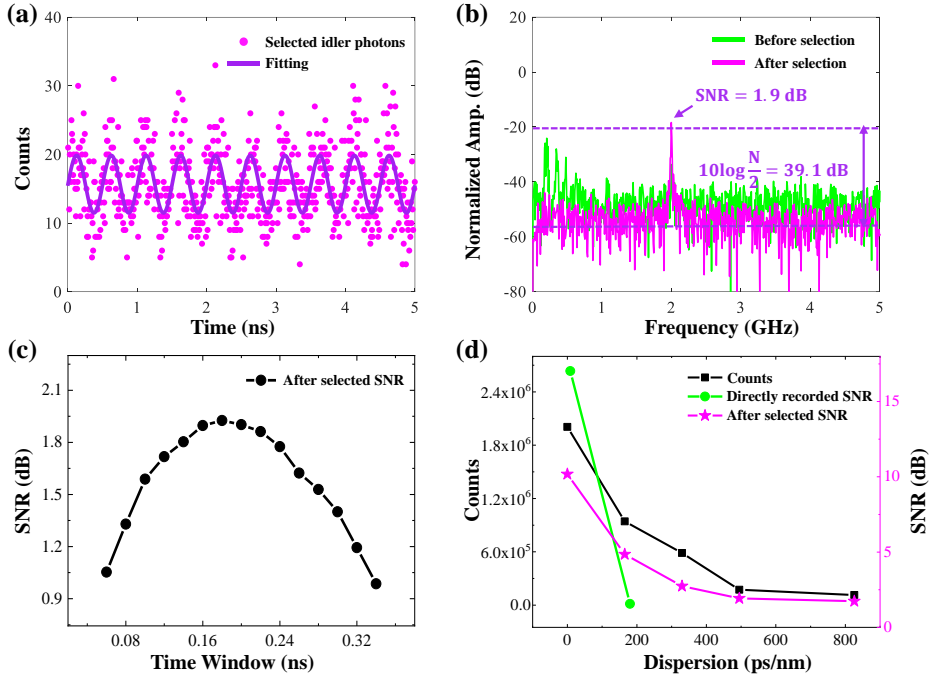

Fig. 6. The experimental results of the distilled RoF signal based on the QMWP scheme, in which the RF signal is set as 2 GHz with a modulation power of 10 dBm and the utilized DCM has a dispersion of 495 ps/nm. (a) The measured idler photon waveform and (b) power spectra after the post-selection. (c) The plot of the distilled RF signal SNR as a function of the heralding time window width, which shows a maximum when the heralding window width is at 0.18 ns. (d) The plot of the RF signal SNR and the corresponding post-selected idler photon number as a function of the dispersion.

## 5. Conclusions

In summary, we have proposed and experimentally demonstrated a QMWP technology by using the time-energy entangled biphoton source as the optical carrier of the RF signal. By combining the single-photon detection with the heralding technique, this QMWP technology not only inherits all the advantages of the SP-MWP[31], it also gains invaluable capability in overcoming the fiber chromatic dispersion distortion associated with ultrashort pulse carriers and the electrical distortion from harmonic. First of all, the QMWP can realize the excellent nonlocal acquisition of the RF signal from the unmodulated optical carrier. Compared with the classical intensity modulation direct detection (IM-DD) technique, the recovered RF signal exhibits not only significantly improved SFDR concerning the second harmonic distortion but excellent resistance to the dispersion distortion associated with ultrashort pulse carriers. Furthermore, the QMWP provides the capability of improving the SFDR of the RF signal on the directly modulated photon carrier and effectively distilling the RF signal from the dispersion distortion. These advantages of QMWP have presented enhancements in ultrafast RoF technology and will open numerous new possibilities in modern communication and networks.

**Funding.**

The National Natural Science Foundation of China (Grant Nos. 12033007, 12103058, 61875205, 61801458, 91836301, 61925505, 61535012, and 61705217), National Key Research and Development Program of China (Grant Nos. 2018YFB2201902, 2018YFB2201901 and 2018YFB2201903), the Frontier Science Key Research Project of Chinese Academy of Sciences (Grant No. QYZDB-SW-SLH007), the Strategic Priority Research Program of CAS (Grant No. XDC07020200), the Youth Innovation Promotion Association, CAS (Grant No. 2021408), the Western Young Scholar Project of CAS (Grant Nos. XAB2019B17 and XAB2019B15), the Chinese Academy of Sciences Key Project (Grant No. ZDRW-KT-2019-1-0103).

**Acknowledgments**

We thank Si Shen and Hao Yu for useful discussions.

**Disclosures.**

The authors declare no conflicts of interest.

**Data availability.**

The data that support the findings of this study are available from the corresponding author upon reasonable request.

**Reference**

1. C. H. Lee, Microwave photonics (CRC press, 2006).
2. K. Xu, R. Wang, Y. Dai, F. Yin, J. Li, Y. Ji, and J. Lin, "Microwave photonics: radio-over-fiber links, systems, and applications," Photonics Res. 2, B54-B63 (2014).
3. T. Kawanishi, "THz and photonic seamless communications," J. Lightwave Technol. 37, 1671-1679 (2019).
4. Y. Yao, F. Zhang, Y. Zhang, X. Ye, D. Zhu, and S. Pan, "Demonstration of ultra-high-resolution photonics-based Kaband inverse synthetic aperture radar imaging," in 2018 Optical Fiber Communications Conference and Exposition (OFC), (IEEE, 2018), 1-3.
5. A. Malacarne, S. Maresca, F. Scotti, B. Hussain, L. Lembo, G. Serafino, A. Bogoni, and P. Ghelfi, "A ultrawide-band VCSEL-based radar-over-fiber system," in 2019 International Topical Meeting on Microwave Photonics (MWP), (IEEE, 2019), 1-4.
6. M. v. Amerongen, "Quantum technologies in defence & security" (2021), retrieved 28 Dec, 2021, https://www.nato.int/docu/review/articles/2021/06/03/quantum-technologies-in-defence-security/index.html.
7. C. Lim and A. Nirmalathas, "Radio-Over-Fiber Technology: Present and Future," J. Lightwave Technol. 39, 881-888 (2021).
8. M. Nakazawa, T. Yamamoto, and K. Tamura, "1.28 Tbit/s-70 km OTDM transmission using third-and fourth-order simultaneous dispersion compensation with a phase modulator," Electron. Lett. 36, 2027-2029 (2000).
9. A. Acń, I. Bloch, H. Buhrman, T. Calarco, C. Eichler, J. Eisert, D. Esteve, N. Gisin, S. J. Glaser, and F. Jelezko, "The quantum technologies roadmap: a European community view," New J. Phys. 20, 080201 (2018).
10. T. B. Pittman, Y. Shih, D. Strekalov, and A. V. Sergienko, "Optical imaging by means of two-photon quantum entanglement," Physical Review A 52, R3429 (1995).
11. D. Strekalov, A. Sergienko, D. Klyshko, and Y. Shih, "Observation of two-photon "ghost" interference and diffraction," Phys. Rev. Lett. 74, 3600 (1995).
12. J. H. Shapiro and R. W. Boyd, "Response to "The physics of ghost imaging—nonlocal interference or local intensity fluctuation correlation?"," Quantum Inf. Process. 11, 1003-1011 (2012).
13. B. E. Saleh, A. F. Abouraddy, A. V. Sergienko, and M. C. Teich, "Duality between partial coherence and partial entanglement," Physical Review A 62, 043816 (2000).

14. A. F. Abouraddy, B. E. Saleh, A. V. Sergienko, and M. C. Teich, "Role of entanglement in two-photon imaging," Phys. Rev. Lett. 87, 123602 (2001).
15. A. F. Abouraddy, B. E. Saleh, A. V. Sergienko, and M. C. Teich, "Entangled-photon Fourier optics," JOSA B 19, 1174-1184 (2002).
16. R. S. Bennink, S. J. Bentley, R. W. Boyd, and J. C. Howell, "Quantum and classical coincidence imaging," Phys. Rev. Lett. 92, 033601 (2004).
17. M. N. O'Sullivan, K. W. C. Chan, and R. W. Boyd, "Comparison of the signal-to-noise characteristics of quantum versus thermal ghost imaging," Physical Review A 82, 053803 (2010).
18. M. Malik, O. S. Magaña-Loaiza, and R. W. Boyd, "Quantum-secured imaging," Appl. Phys. Lett. 101, 241103 (2012).
19. E. Lopaeva, I. R. Berchera, I. P. Degiovanni, S. Olivares, G. Brida, and M. Genovese, "Experimental realization of quantum illumination," Phys. Rev. Lett. 110, 153603 (2013).
20. N. Samantaray, I. Ruo-Berchera, A. Meda, and M. Genovese, "Realization of the first sub-shot-noise wide field microscope," Light Sci. Appl. 6, e17005-e17005 (2017).
21. R. Quan, Y. Zhai, M. Wang, F. Hou, S. Wang, X. Xiang, T. Liu, S. Zhang, and R. Dong, "Demonstration of quantum synchronization based on second-order quantum coherence of entangled photons," Sci. Rep. 6, 1-8 (2016).
22. F. Hou, R. Quan, R. Dong, X. Xiang, B. Li, T. Liu, X. Yang, H. Li, L. You, and Z. Wang, "Fiber-optic two-way quantum time transfer with frequency-entangled pulses," Physical Review A 100, 023849 (2019).
23. R. Quan, R. Dong, Y. Zhai, F. Hou, X. Xiang, H. Zhou, C. Lv, Z. Wang, L. You, and T. Liu, "Simulation and realization of a second-order quantum-interference-based quantum clock synchronization at the femtosecond level," Opt. Lett. 44, 614-617 (2019).
24. Y. Liu, R. Quan, X. Xiang, H. Hong, M. Cao, T. Liu, R. Dong, and S. Zhang, "Quantum clock synchronization over 20-km multiple segmented fibers with frequency-correlated photon pairs and HOM interference," Appl. Phys. Lett. 119, 144003 (2021).
25. J. Nunn, L. Wright, C. Söller, L. Zhang, I. Walmsley, and B. Smith, "Large-alphabet time-frequency entangled quantum key distribution by means of time-to-frequency conversion," Opt. Express 21, 15959-15973 (2013).
26. J. M. Lukens, A. Dezfooliyan, C. Langrock, M. M. Fejer, D. E. Leaird, and A. M. Weiner, "Orthogonal spectral coding of entangled photons," Phys. Rev. Lett. 112, 133602 (2014).
27. A. Yabushita and T. Kobayashi, "Spectroscopy by frequency-entangled photon pairs," Physical Review A 69, 013806 (2004).
28. R. Whittaker, C. Erven, A. Neville, M. Berry, J. O'Brien, H. Cable, and J. Matthews, "Absorption spectroscopy at the ultimate quantum limit from single-photon states," New J. Phys. 19, 023013 (2017).
29. V. Averchenko, D. Sych, G. Schunk, U. Vogl, C. Marquardt, and G. Leuchs, "Temporal shaping of single photons enabled by entanglement," Physical Review A 96, 043822 (2017).
30. V. Averchenko, D. Sych, C. Marquardt, and G. Leuchs, "Efficient generation of temporally shaped photons using nonlocal spectral filtering," Physical Review A 101, 013808 (2020).
31. Y. Yang, Y. Jin, X. Xiang, T. Hao, W. Li, T. Liu, S. Zhang, N. Zhu, R. Dong, and M. Li, "Single-photon microwave photonics," Science Bulletin (2021).
32. Y. Zhang, F. Hou, T. Liu, X. Zhang, S. Zhang, and R. Dong, " Generation and quantum characterization of miniaturized frequency entangled source in telecommunication band based on type-II periodically poled lithium niobate waveguide," Acta Physica Sinica 67, 144204 (2018).
33. J. Wu, L. You, S. Chen, H. Li, Y. He, C. Lv, Z. Wang, and X. Xie, "Improving the timing jitter of a superconducting nanowire single-photon detection system," Appl. Opt. 56, 2195-2200 (2017).
34. M. Frigo and S. G. Johnson, "FFTW: An adaptive software architecture for the FFT," in Proceedings of the 1998 IEEE International Conference on Acoustics, Speech and Signal Processing, ICASSP'98 (Cat. No. 98CH36181), (IEEE, 1998), 1381-1384.